\documentclass[12pt]{article}

\usepackage{amssymb,amsmath}

\usepackage{pb-diagram,lamsarrow,pb-lams}

\DeclareMathAlphabet{\bb}{U}{msb}{m}{n}
\gdef\C{\bb C}
\gdef\dZ{\bb Z}

\gdef\dS{\bb S}
\gdef\R{\bb R}
\gdef\K{\bb K}
\gdef\BH{\bb H}
\gdef\F{\bb F}

\DeclareMathOperator{\End}{End}
\DeclareMathOperator{\spin}{{\bf Spin}}
\DeclareMathOperator{\pin}{{\bf Pin}}

\DeclareMathOperator{\Id}{Id}

\DeclareMathOperator{\Aut}{Aut}
\DeclareMathOperator{\sAut}{{\sf Aut}}
\DeclareMathOperator{\sExt}{{\sf Ext}}

\DeclareMathOperator{\Ext}{Ext}

\newcommand{\cA}{\mathcal{A}}

\newcommand{\M}{{\bf\sf M}}

\newcommand{\sA}{{\sf A}}

\newcommand{\sI}{{\sf I}}
\newcommand{\sW}{{\sf W}}
\newcommand{\sE}{{\sf E}}
\newcommand{\sC}{{\sf C}}
\newcommand{\sF}{{\sf F}}
\newcommand{\sT}{{\sf T}}
\newcommand{\sS}{{\sf S}}

\newcommand{\sK}{{\sf K}}

\newcommand{\bx}{{\bf x}}

\newcommand{\bZ}{{\bf Z}}

\newcommand{\Lip}{\boldsymbol{\Gamma}}

\newcommand{\cl}{C\kern -0.2em \ell}

\newcommand{\e}{\mbox{\bf e}}

\title{The CPT Group in the de Sitter Space}

\author{V. V. Varlamov\\
{\small\it Department of Mathematics, Siberia State University of Industry,}\\
{\small\it Kirova 42, Novokuznetsk 654007, Russia}}
\date{}

\begin{document}

\maketitle

\begin{abstract}

$P$-, $T$-, $C$-transformations of the Dirac field in the de Sitter
space are studied in the framework of an automorphism set of
Clifford algebras. Finite group structure of the discrete
transformations is elucidated. It is shown that $CPT$ groups of the
Dirac field in Minkowski and de Sitter spaces are isomorphic.

\end{abstract}


\section{Introduction}
In 1935, Dirac \cite{Dir35} introduced relativistic wave equations in a
five-dimensional pseudoeuclidean space (de Sitter space),
\begin{equation}\label{DirSit}
(i\gamma_0\partial_0+i\gamma_k\partial_k-m)\psi=0
\end{equation}
or
\[
(i\gamma_\mu\partial_\mu+m)\psi=0,
\]
where five $4\times 4$ Dirac matrices $\gamma_\mu$ satisfy the relations
\[
\gamma_\mu\gamma_\nu+\gamma_\nu\gamma_\mu=2g_{\mu\nu},\quad
\mu=0,1,2,3,4.
\]
Later on, Fushchych and Krivsky \cite{FK69} showed that equations
(\ref{DirSit}) do not describe particles and antiparticles symmetrically.
This result means that the equations (\ref{DirSit}) are non-invariant
under action of transformations of the type $CPT$.

In general,
usual practice of definition of the discrete symmetries from the analysis of
relativistic wave equations
does not give a complete and consistent  theory of the discrete
transformations. In the standard approach, except a well studied case of the
spin $j=1/2$ (Dirac equations), a situation with the discrete symmetries
remains unclear for the fields of higher spin $j>1/2$. Moreover, Lee and
Wick \cite{LW66} claimed that ``the situation is clearly an
unsatisfactory one from a fundamental point of view". It is obvious that
a main reason of this is an absence of a fully adequate formalism for
description of higher-spin fields (all widely accepted higher-spin
formalisms have many intrinsic contradictions and difficulties).

Taking into account the present status of higher-spin theories, we must
construct an alternative approach for definition of the discrete
transformations without handling to analysis of relativistic wave equations.
The purely algebraic approach based on the automorphism set of Clifford
algebras has been proposed in the works \cite{Var99,Var00,Var041}.

Following to \cite{Var041}, in the present paper we study $P$-, $T$-, $C$-
transformations for the Dirac field in the de Sitter space $\R^{4,1}$.
In the section 2 we give some basic facts concerning a relationship
between finite groups and Clifford algebras. An universal covering of the
de Sitter group is given in the section 3. $CPT$-groups of the Dirac
field in the Minkowski spacetime and their isomorphisms to finite groups
are studied in the section 4. In the section 5 we define a $CPT$ group
in the space $\R^{4,1}$.

\section{The Dirac group}
As is known \cite{Sal81a,Sal82,Sal84,Bra85,Sha94}, a structure of the
Clifford algebras admits a very elegant description in terms of finite
groups.
In accordance with a multiplication rule
\begin{equation}\label{e1}
\e^2_i=\sigma(p-i)\e_0,\quad\e_i\e_j=-\e_j\e_i,
\end{equation}
\begin{equation}\label{e2}
\sigma(n)=\left\{\begin{array}{rl}
-1 & \mbox{if $n\leq 0$},\\
+1 & \mbox{if $n>0$},
\end{array}\right.
\end{equation}
basis elements  of the Clifford algebra
$\cl_{p,q}$ (the algebra over the field of real numbers,
$\F=\R$) form a finite group of order
$2^{n+1}$,
\begin{equation}\label{FG}
G(p,q)=\left\{\pm 1,\,\pm\e_i,\,\pm\e_i\e_j,\,\pm\e_i\e_j\e_k,\,\ldots,\,
\pm\e_1\e_2\cdots\e_n\right\}\quad(i<j<\ldots).
\end{equation}
The Dirac group is a particular case of (\ref{FG}).

In turn, the Dirac algebra $\C_4$ is a complexification of {\it the
spacetime algebra} $\cl_{1,3}$, $\C_4=\C\otimes\cl_{1,3}$.
An arbitrary element of $\C_4$ has the form
\begin{equation}\label{ArbD}
\cA=a^0\e_0+\sum^4_{i=1}a^i\e_i+\sum^4_{i=1}\sum^4_{j=1}a^{ij}\e_i\e_j
+\sum^4_{i=1}\sum^4_{j=1}\sum^4_{k=1}a^{ijk}\e_i\e_j\e_k+
a^{1234}\e_1\e_2\e_3\e_4,
\end{equation}
where the coefficients $a^0$, $a^i$, $a^{ij}$, $a^{ijk}$, $a^{1234}$
are complex numbers.

If we consider a spinor representation (the left regular representation
in a spinspace $\dS_4$), then the units $\e_i$ of the algebra $\C_4$ are replaced
by $\gamma$-matrices via the rule $\gamma_i=\gamma(\e_i)$, where
$\gamma$ is a mapping of the form
$\cl_{p,q}\overset{\gamma}{\longrightarrow}\End_{\K}(\dS)$.
With the physical purposes
we choose from the set of all isomorphic spinbasises
of $\C_4$ a so called
{\it canonical basis}
\begin{equation}\label{CB}
\gamma^C_0=\begin{pmatrix}
-\boldsymbol{1}_2 & 0\\
0 & \boldsymbol{1}_2
\end{pmatrix},\quad
\gamma^C_k=\begin{pmatrix}
0 & \sigma_k\\
-\sigma_k & 0
\end{pmatrix}
\end{equation}
or the {\it Weyl basis}
\begin{equation}\label{WB}
\gamma^W_m=\begin{pmatrix}
0 & \sigma_m\\
\overline{\sigma}_m & 0
\end{pmatrix},
\end{equation}
which related with (\ref{CB}) by the following similarity transformation:
\[
\Gamma^W=X\Gamma^CX^{-1},\quad X=\frac{1}{\sqrt{2}}\begin{pmatrix}
\boldsymbol{1}_2 & -\boldsymbol{1}_2\\
\boldsymbol{1}_2 & \boldsymbol{1}_2
\end{pmatrix}.
\]
It is of interest to consider a {\it Majorana basis}
\[
\gamma^M_0=\begin{pmatrix}
0 & -\sigma_2\\
-\sigma_2 & 0
\end{pmatrix},\quad
\gamma^M_1=\begin{pmatrix}
0 & i\sigma_3\\
i\sigma_3 & 0
\end{pmatrix},
\]
\begin{equation}\label{MB}
\gamma^M_2=\begin{pmatrix}
i\boldsymbol{1}_2 & 0\\
0 & -i\boldsymbol{1}_2
\end{pmatrix},\quad
\gamma^M_3=\begin{pmatrix}
0 & -i\sigma_1\\
-i\sigma_1 & 0
\end{pmatrix},
\end{equation}
which, in turn, related with the Weyl basis by a similarity transformation
of the form
\[
\Gamma^W=Y\Gamma^MY^{-1},\quad Y=\frac{1}{\sqrt{2}}\begin{pmatrix}
\boldsymbol{1}_2 & i\boldsymbol{1}_2\\
\varepsilon\boldsymbol{1}_2 & -i\varepsilon\boldsymbol{1}_2
\end{pmatrix}.
\]
All the basises
(\ref{CB}), (\ref{WB}) and
(\ref{MB}) are isomorphic within the algebra $\C_4$. Moreover, these three
basises are isomorphic within the real subalgebra
$\cl_{1,3}\subset\C_4$.

In dependence on a division ring structure $\K$, the Dirac algebra has
five real subalgebras.
Three subalgebras with the quaternion ring
$\K\simeq\BH$: the spacetime algebra $\cl_{1,3}$, $\cl_{4,0}$ and
$\cl_{0,4}$. Two subalgebras with the real ring $\K\simeq\R$: the
{\it Majorana} $\cl_{3,1}$ and {\it Klein} $\cl_{2,2}$ algebras.
In accordance with (\ref{FG}) the each real subalgebra
$\cl_{p,q}\subset\C_4$ induces a finite group $G(p,q)$.
Let us consider in detail the structure of these finite groups.

We can work in any basises (\ref{CB})--(\ref{MB}).
Owing to (\ref{FG}) a
{\it spacetime group} is defined by the followng set:
\begin{multline}
G(1,3)=\{\pm 1,\,\pm\gamma_0,\,\pm\gamma_1,\,\pm\gamma_2,\,
\pm\gamma_3,\,\pm\gamma_0\gamma_1,\,
\pm\gamma_0\gamma_2,\\
\pm\gamma_0\gamma_3,\,
\pm\gamma_1\gamma_2,\,\pm\gamma_1\gamma_3,\,
\pm\gamma_2\gamma_3,\,\pm\gamma_0\gamma_1\gamma_2,\,
\pm\gamma_0\gamma_1\gamma_3,\,\\
\pm\gamma_0\gamma_2\gamma_3,\,\pm\gamma_1\gamma_2\gamma_3,\,
\pm\gamma_0\gamma_1\gamma_2\gamma_3\}.\label{DG}
\end{multline}
It is a finite group of order 32\footnote{We can write out
the multiplication table of this group (a Cayley table). However,
an explicit form of the Cayley table for the group
$G(1,3)$ is very cumbersome; this table contains $2^{10}$ cells.}
with an order structure
$(11,20)$. Moreover,
$G(1,3)$ is an {\it extraspecial two-group} \cite{Sal84,Bra85}.
In Salingaros notation the following isomorphism holds:
\[
G(1,3)=N_4\simeq Q_4\circ D_4,
\]
where $Q_4$ is a quaternion group, $D_4$ is a dihedral group,
$\circ$ is a {\it central product} ($Q_4$ and $D_4$ are finite groups of
order 8). A center of the group $G(1,3)$ is isomorphic to a cyclic group
$\dZ_2$. $G(1,3)$ is non-Abelian group which contains many subgroups both
Abelian and non-Abelian.
For example, the group of fundamental automorphisms of the algebra
$\cl_{1,3}$ \cite{Var00} is an Abelian subgroup of
$G(1,3)$, $\Aut(\cl_{1,3})=\{\Id,\star,
\widetilde{\phantom{cc}},\widetilde{\star}\}\simeq\{1,P,T,PT\}
\simeq\dZ_2\otimes\dZ_2
\subset G(1,3)$, where $\dZ_2$ is the Gauss-Klein group. In turn,
an extended automorphism group of the algebra
$\cl_{1,3}$ is a non-Abelian subgroup of
$G(1,3)$, $\Ext(\cl_{1,3})\simeq
\overset{\ast}{\dZ}_4\otimes\dZ_2\subset G(1,3)$ \cite{Var041,Var042}.

Finite groups $G(4,0)$ and $G(0,4)$, corresponding to the subalgebras
$\cl_{4,0}$ and $\cl_{0,4}$, are isomorphic to each
other\footnote{This group isomorphism is a direct consequence of the algebra
isomorphism
$\cl_{4,0}\simeq\cl_{0,4}$.}, since these groups possess the order structure
$(11,20)$. Therefore,
\[
G(4,0)\simeq G(0,4)=N_4\simeq Q_4\circ D_4.
\]

Let us consider now the {\it Majorana group} $G(3,1)$. Since the subalgebra
$\cl_{3,1}$ has a real division ring $\K\simeq\R$, then none of the
spinbasises (\ref{CB}), (\ref{WB}), (\ref{MB}) can be used in this case
(all these basises are defined over the ring $\K\simeq\BH$).
One of the permissible real spinbasises of the algebra
$\cl_{3,1}$ has the form\footnote{In fact, a definition procedure of the
spinbasis over the ring $\K$ is hardly fixed and depends on the structure
of primitive idempotents $f$ of the algebra $\cl_{p,q}$.
The primitive idempotent for the algebra
$\cl_{3,1}$ is defined by an expression
$f=\frac{1}{4}(1+\e_1)(1+\e_{34})$, and a division ring has the form
$\K=f\cl_{3,1}f=\{1\}\simeq\R$.}:
\[
\gamma_0=\begin{pmatrix}
1 & 0 & 0 & 0\\
0 &-1 & 0 & 0\\
0 & 0 &-1 & 0\\
0 & 0 & 0 & 1
\end{pmatrix},\quad\gamma_1=\begin{pmatrix}
0 & 1 & 0 & 0\\
1 & 0 & 0 & 0\\
0 & 0 & 0 & 1\\
0 & 0 & 1 & 0
\end{pmatrix},
\]
\[
\gamma_2=\begin{pmatrix}
0 & 0 & 1 & 0\\
0 & 0 & 0 &-1\\
1 & 0 & 0 & 0\\
0 & -1 & 0 & 0
\end{pmatrix},\quad\gamma_3=\begin{pmatrix}
0 & 0 & -1 & 0\\
0 & 0 & 0 & 1\\
1 & 0 & 0 & 0\\
0 & -1 & 0 & 0
\end{pmatrix}.
\]
The Majorana group $G(3,1)$ with the order structure $(19,12)$ is a central
product of the two groups $D_4$:
\[
G(3,1)=N_3\simeq N_1\circ N_1\simeq D_4\circ D_4.
\]
$G(3,1)$ is the non-Abelian group; a center of the group is isomorphic to
$\dZ_2$.

The same isomorphism takes place for the {\it Klein group}:
\[
G(2,2)=N_3\simeq D_4\circ D_4.
\]

Thus, the real subalgebras of the Dirac algebra $\C_4$ form five finite
groups of order 32. Three subalgebras with the ring
$\K\simeq\BH$ form finite groups which isomorphic to the central product
$Q_4\circ D_4$, and two subalgebras with the ring $\K\simeq\R$ form finite
groups defined by the product $D_4\circ D_4$.

Let us define now a finite group which is immediately related with the Dirac
algebra $\C_4$. With this end in view it is necessary to consider the
following isomorphism: $\C_4\simeq\cl_{4,1}$, where $\cl_{4,1}$ is the
Clifford algebra over the field $\F=\R$ with a complex division ring
$\K\simeq\C$. An arbitrary element of $\cl_{4,1}$ has the form
\begin{multline}
\cA=b^0\e_0+\sum^5_{i=1}b^i\e_i+\sum^5_{i=1}\sum^5_{j=1}b^{ij}\e_i\e_j+
\sum^5_{i=1}\sum^5_{j=1}\sum^5_{k=1}b^{ijk}\e_i\e_j\e_k+\\
+\sum^5_{i=1}\sum^5_{j=1}\sum^5_{k=1}\sum^5_{l=1}b^{ijkl}
\e_i\e_j\e_k\e_l+b^{12345}\e_1\e_2\e_3\e_4\e_5,\label{Arb}
\end{multline}
where the coefficients $b^0,b^{ij},\ldots$ are real numbers. It is easy to
verify that a volume element
$\omega=\e_{12345}=\e_1\e_2\e_3\e_4\e_5$ commutes with all other
elements of the algebra $\cl_{4,1}$, that is,
$\omega=\e_{12345}$ belongs to a center $\bZ_{4,1}$ of
$\cl_{4,1}$. A general definition of the center $\bZ_{p,q}$ of $\cl_{p,q}$
is
\[
\bZ_{p,q}=\begin{cases}
\{1\},& \text{if $p-q\equiv 0,2,4,6\pmod{8}$};\\
\{1,\,\omega\},& \text{if $p-q\equiv 1,3,5,7\pmod{8}$}.
\end{cases}
\]
This property of the element $\omega$ allows us to rewrite the arbitrary
element (\ref{Arb}) in the following form:
\begin{multline}
\cA=(b^0+\omega b^{12345})\e_0+(b^1+\omega b^{2345})\e_1+
(b^2+\omega b^{1345})\e_2+(b^3+\omega b^{1245})\e_3+\\
+(b^4+\omega b^{1235})\e_4+(b^{12}+\omega b^{345})\e_{12}+
(b^{13}+\omega b^{245})\e_{13}+(b^{14}+\omega b^{235})\e_{14}+\\
+(b^{23}+\omega b^{145})\e_{23}+(b^{24}+\omega b^{135})\e_{24}+
(b^{34}+\omega b^{125})\e_{34}+(b^{123}+\omega b^{45})\e_{123}+\\
+(b^{124}+\omega b^{35})\e_{124}+(b^{234}+\omega b^{15})\e_{234}+
(b^{134}+\omega b^{25})\e_{134}+(b^{1234}+\omega b^5)\e_{1234}.
\label{Arb2}
\end{multline}
Moreover, a square of the element $\omega$ is equal to $-1$.
A general definition of the square of
$\omega$ is
\[
\omega^2=\begin{cases}
-1, & \text{if $p-q\equiv 2,3,6,7\pmod{8}$};\\
+1, & \text{if $p-q\equiv 0,1,4,5\pmod{8}$}.
\end{cases}
\]
Therefore, the element $\omega$ can be identified with an imaginary unit,
$\omega\equiv i$. Hence it follows that expressions, standing in the
brackets in (\ref{Arb2}), are complex numbers of the form
$a^0=b^0+ib^{12345}$,
$a^1=b^1+ib^{2345}$, $\ldots$, $a^{1234}=b^{1234}+ib^5$.
Thus, the element (\ref{Arb2}) coincides with the arbitrary element
(\ref{ArbD}) of $\C_4$; it proves the isomorphism
$\C_4\simeq\cl_{4,1}$.

So, in accordance with (\ref{FG}) the
{\it Dirac group}
is defined by the following set:
\begin{multline}
G(4,1)=\{\pm1,\,\pm\e_1,\,\pm\e_2,\,\pm\e_3,\,\pm\e_4,\,\pm\e_5,\,
\pm\e_{12},\,\pm\e_{13},\,
\pm\e_{14},\\
\pm\e_{15},\,\pm\e_{23},\,\pm\e_{24},\,
\pm\e_{25},\,\pm\e_{34},\,\pm\e_{35},\,\pm\e_{45},\,
\pm\e_{123},\,\pm\e_{124},\\
\pm\e_{125},\,\pm\e_{134},\,\pm\e_{135},\,\pm\e_{145},\,
\pm\e_{234},\,\pm\e_{235},\,\pm\e_{245},\,
\pm\e_{345},\\
\pm\e_{1234},\,
\pm\e_{1235},\,\pm\e_{1245},\,\pm\e_{1345},\,
\pm\e_{2345},\,\pm\e_{12345}\}.
\label{DirG}
\end{multline}
It is a finite group of order 64 with the order structure $(31,32)$.
The Cayley table of $G(4,1)$ consists of $2^{12}$ cells. In common with all
finite groups considered previously, the Dirac group is an extraspecial
two-group. For this group the following isomorphism holds:
\[
G(4,1)=S_2\simeq N_4\circ\dZ_4\simeq Q_4\circ D_4\circ\dZ_4.
\]
The center of $G(4,1)$ is isomorphic to a complex group $\dZ_4$.
$G(4,1)$ is the non-Abelian group (as all Salingaros groups, except the first
three groups $\dZ_2$, $\Omega_0=\dZ_2\otimes\dZ_2$ and
$S_0=\dZ_4$).

As is known, a five-dimensional pseudoeuclidean space $\R^{4,1}$
(so called {\it de Sitter space}) is associated with the algebra
$\cl_{4,1}$.
It is interesting to note that the
{\it anti-de Sitter space} $\R^{3,2}$, associated with the algebra
$\cl_{3,2}$, leads to the following extraspecial group of order 64:
\[
G(3,2)=\Omega_3\simeq N_3\circ D_2\simeq D_4\circ D_4\circ D_2.
\]
\subsection{Spinor representation of the Dirac group}
In virtue of the isomorphism $\cl_{4,1}\simeq\C_4\simeq\M_4(\C)$ for the
algebra $\cl_{4,1}$ there exists a spinor representation within the full
matrix algebra $\M_4(\C)$ over the field $\F=\C$.
Let us find spinor representations of the units
$\e_i$ ($i=1,\ldots,5$) of the algebra $\cl_{4,1}$, that is,
$\gamma_i=\gamma(\e_i)$. To this end it is more convenient to use
{\it a Brauer-Weyl representation} \cite{BW35}. This representation
is defined by the following tensor products of
$m$ Pauli matrices:
\begin{equation}\label{6.6}
{\renewcommand{\arraystretch}{1.2}
\begin{array}{lcl}
\gamma_{1}&=&\sigma_{1}\otimes\boldsymbol{1}_2\otimes\cdots\otimes
\boldsymbol{1}_2\otimes
\boldsymbol{1}_2\otimes\boldsymbol{1}_2,\\
\gamma_{2}&=&\sigma_{3}\otimes\sigma_{1}\otimes\boldsymbol{1}_2\otimes\cdots\otimes
\boldsymbol{1}_2\otimes\boldsymbol{1}_2,\\
\gamma_{3}&=&\sigma_{3}\otimes\sigma_{3}\otimes\sigma_{1}\otimes\boldsymbol{1}_2
\otimes\cdots\otimes\boldsymbol{1}_2,\\
\hdotsfor[2]{3}\\
\gamma_{m}&=&\sigma_{3}\otimes\sigma_{3}\otimes\cdots\otimes\sigma_{3}
\otimes\sigma_{1},\\
\gamma_{m+1}&=&\sigma_{2}\otimes\boldsymbol{1}_2\otimes\cdots\otimes\boldsymbol{1}_2,\\
\gamma_{m+2}&=&\sigma_{3}\otimes\sigma_{2}\otimes\boldsymbol{1}_2\otimes\cdots\otimes
\boldsymbol{1}_2,\\
\hdotsfor[2]{3}\\
\gamma_{2m}&=&\sigma_{3}\otimes\sigma_{3}\otimes\cdots\otimes\sigma_{3}
\otimes\sigma_{2}.
\end{array}}\end{equation}
In general case, this representation defines a spinbasis of the
even-dimensional Clifford algebra
$\cl_{p,q}$ $(p+q=2m\equiv 0\pmod{2})$ over the ring
$\K\simeq\BH$. The case $\K\simeq\R$ is easy realized via the replacement
of $\sigma_i$ by the real matrices.

In the case of $n=2m+1$ we add to the tensor products (\ref{6.6})
the following matrix:
\begin{equation}\label{6.12}
\gamma_{2m+1}=\underbrace{\sigma_{3}\otimes\sigma_{3}\otimes\cdots\otimes\sigma_{3}
}_{m\;\text{times}},
\end{equation}
It is easy to verify that this matrix satisfies the conditions
\begin{gather}
\gamma^{2}_{2m+1}=\boldsymbol{1}_m,
\quad \gamma_{2m+1}\gamma_{i}=-\gamma_{i}\gamma_{2m+1},\nonumber\\
i=1,2,\ldots, m.\nonumber
\end{gather}
The product
$\gamma_{1}\gamma_{2}\cdots\gamma_{2m}\gamma_{2m+1}$ commutes with all the
products of the form
$\gamma^{\alpha_1}_1\gamma^{\alpha_2}_2\cdots\gamma^{\alpha_{2m}}_{2m}$,
where $\alpha_1,\alpha_2,\ldots,\alpha_{2m}=0,1$.

In the Brauer-Weyl representation a spinbasis of the 32-dimensional
algebra $\cl_{4,1}$ is defined as follows
\[
\gamma_1=\sigma_1\otimes\boldsymbol{1}_2=\begin{pmatrix}
0 & \boldsymbol{1}_2\\
\boldsymbol{1}_2 & 0
\end{pmatrix},\quad
\gamma_2=\sigma_3\otimes\sigma_1=\begin{pmatrix}
\sigma_1 & 0\\
0 & -\sigma_1
\end{pmatrix},
\]
\[
\gamma_3=\sigma_2\otimes\boldsymbol{1}_2=\begin{pmatrix}
0 & -i\boldsymbol{1}_2\\
i\boldsymbol{1}_2 & 0
\end{pmatrix},\quad
\gamma_4=\sigma_3\otimes\sigma_2=\begin{pmatrix}
\sigma_2 & 0\\
0 & -\sigma_2
\end{pmatrix},
\]
\begin{equation}\label{SS}
\gamma_5=i\sigma_3\otimes\sigma_3=\begin{pmatrix}
i\sigma_3 & 0\\
0 & -i\sigma_3
\end{pmatrix}.
\end{equation}
It is easy to verify that the product
$\gamma_1\gamma_2\gamma_3\gamma_4\gamma_5$ commutes with all the basis
elements of the algebra $\cl_{4,1}\simeq\M_4(\C)$.

Thus, a spinor representation of the Dirac group has the form
\begin{equation}\label{SD}
G(4,1)=\{\pm 1,\pm\gamma_1,\pm\gamma_2,\ldots,\pm\gamma_{12345}\}.
\end{equation}
In common with (\ref{DirG}) the group (\ref{SD}) is the extraspecial finite
group of order 64 and is isomorphic to the central product
$Q_4\circ D_4\circ\dZ_4$.

\section{The de Sitter group}
First of all, let us give several definitions which will be used below.
The Lipschitz group $\Lip_{p,q}$, also called the Clifford group,
introduced by Lipschitz in 1886
(Lipschitz used the Clifford algebras for the study of rotation groups in
multidimensional spaces),
may be defined as the subgroup of invertible elements $s$ of the algebra
$\cl_{p,q}$:
\begin{eqnarray}
\Lip_{p,q}&=&\left\{s\in\cl_{p,q}\;|\;\forall{\bf x}\in\R^{p,q},\,s{\bf x}s^{-1}
\in\R^{p,q}\right\},\nonumber\\
\Lip_{p,q}&=&\left\{s\in\cl^+_{p,q}\cup\cl^-_{p,q}\;|\;\forall {\bf x}\in\R^{p,q},
\,s{\bf x}s^{-1}\in\R^{p,q}\right\}.\nonumber
\end{eqnarray}\begin{sloppypar}\noindent
The set $\Lip^+_{p,q}=\Lip_{p,q}\cap\cl^+_{p,q}$ is called {\it
special Lipschitz group} \cite{Che54}. Let $N:\;\cl_{p,q}
\rightarrow\cl_{p,q},\;N(x)=x\widetilde{x}$. If $\bx\in\R^{p,q}$, then
$N(\bx)=\bx(-\bx)=-\bx^2=-Q(\bx)$. Further, the group $\Lip_{p,q}$ has a
subgroup\end{sloppypar}
\begin{equation}\label{Pin}
\pin(p,q)=\left\{s\in\Lip_{p,q}\;|\;N(s)=\pm 1\right\}.
\end{equation}
It is easy to see that $\pin(p,q)\simeq\Lip_{p,q}/\R^\ast_+$,
where $\R^\ast_+$ is a set of non-negative real numbers,
$\R^\ast=\R-\{0\}$. Analogously,
{\it a spinor group} $\spin(p,q)$ is defined by the set
\begin{equation}\label{Spin}
\spin(p,q)=\left\{s\in\Lip^+_{p,q}\;|\;N(s)=\pm 1\right\}.
\end{equation}
It is obvious that
$
\spin(p,q)=\pin(p,q)\cap\cl^+_{p,q}
$
and $\spin(p,q)\simeq\Lip^+_{p,q}/\R^\ast_+$.
The group $\spin(p,q)$ contains a subgroup
\begin{equation}\label{Spin+}
\spin_+(p,q)=\left\{s\in\spin(p,q)|N(s)=1\right\}.
\end{equation}\begin{sloppypar}\noindent
The orthogonal group $O(p,q)$, special orthogonal group
$SO(p,q)$ and special orthogonal group $SO_+(p,q)$ with the unit
determinant, are isomorphic correspondingly to the following quotient groups:
\end{sloppypar}
\begin{eqnarray}
O(p,q)&\simeq&\pin(p,q)/\dZ_2,\nonumber\\
SO(p,q)&\simeq&\spin(p,q)/\dZ_2,\nonumber\\
SO_+(p,q)&\simeq&\spin_+(p,q)/\dZ_2,\nonumber
\end{eqnarray}
where the kernel is $\dZ_2=\left\{+1,-1\right\}$. Thus, the groups $\pin(p,q)$,
$\spin(p,q)$ and $\spin_+(p,q)$ are universal coverings of the groups $O(p,q)$,
$SO(p,q)$ and $SO_+(p,q)$, respectively.

Further, since $\cl^+_{p,q}\simeq\cl^+_{q,p}$, then
\[
\spin(p,q)\simeq\spin(q,p).
\]
In contrast to this, the groups $\pin(p,q)$ and $\pin(q,p)$ are non-isomorphic.

So, in our case the Clifford-Lipschitz group, corresponding the de Sitter
space $\R^{4,1}$, has the form
\[
\Lip_{4,1}=\left\{s\in\cl^+_{4,1}\cup\cl^-_{4,1}\;|\;\forall {\bf x}\in\R^{4,1},
\,s{\bf x}s^{-1}\in\R^{4,1}\right\}.
\]
The special Clifford-Lipschitz group for $\R^{4,1}$ is defined as
\[
\Lip^+_{4,1}=\Lip_{4,1}\cap\cl^+_{4,1},
\]
at this point there is an isomorphism $\cl^+_{4,1}\simeq\cl_{1,3}$, that is,
the subalgebra $\cl^+_{4,1}$ of all even elements of the algebra
$\cl_{4,1}$ is isomorphic to the spacetime algebra $\cl_{1,3}$. Further,
\begin{eqnarray}
\pin(4,1)&=&\{s\in\Lip_{4,1}|N(s)=\pm 1\},\nonumber\\
\spin(4,1)&=&\{s\in\Lip^+_{4,1}|N(s)=\pm 1\},\nonumber
\end{eqnarray}
at this point, in virtue of $\cl^+_{4,1}\simeq\cl_{1,3}$, we have
\begin{equation}\label{DeS}
\spin(4,1)\simeq\pin(1,3).
\end{equation}
The group $\pin(4,1)$ is {\it an universal covering} of the de Sitter
group $O(4,1)$:
\[
O(4,1)\simeq\pin(4,1)/\dZ_2.
\]\begin{sloppypar}\noindent
From (\ref{DeS}) it follows that a rotation group
$SO(4,1)\simeq\spin(4,1)/\dZ_2$ of the space $\R^{4,1}$ is isomorphic
to the general Lorentz group $O(1,3)\simeq\pin(1,3)/\dZ_2$.
Defining the general Lorentz group as a semidirect product
$O(1,3)=O_+(1,3)\odot\{1,P,T,PT\}$, where
$O_+(1,3)\simeq\spin_+(1,3)/\dZ_2\simeq SL(2,\C)$ is a connected component
of the Lorentz group, we can express the universal covering
$\pin(4,1)$ via the Shirokov-D\c{a}browski
$PT$-structures \cite{Shi60,Dab88} or via more general
$CPT$-structures \cite{Var041,Var042}.
\end{sloppypar}\begin{sloppypar}
The relation between the Dirac algebra $\C_4\simeq\cl_{4,1}$, Dirac group
$G(4,1)$, Clifford-Lipschitz groups $\Lip_{4,1}$ and $\Lip^+_{4,1}$,
universal coverings $\pin(4,1)$ and $\spin(4,1)$, de Sitter
$O(4,1)$ and Lorentz $O(1,3)$ groups, can be represented by the following
scheme:\end{sloppypar}
\[
\dgARROWLENGTH=2.5em
\begin{diagram}
\node{\C_4\simeq\cl_{4,1}}\arrow{s,<>}\arrow{e}\node{\Lip_{4,1}}
\arrow{s}\arrow{e}\node{\pin(4,1)}\arrow{e}\node{O(4,1)}\\
\node{G(4,1)}\node{\Lip^+_{4,1}}\arrow{e}\node{\spin(4,1)}\arrow{s}
\arrow{e}\node{SO(4,1)}\arrow{s,<>}\\
\node[3]{\pin(1,3)}\arrow{e}\node{O(1,3)}
\end{diagram}
\]
Analogous schemes can be defined for the real subalgebras
(correspondingly, finite groups) $\cl_{p,q}\leftrightarrow G(p,q)$ of the
Dirac algebra (group) $\C_4\leftrightarrow G(4,1)$, $p+q=4$.

\section{The $CPT$-group in the Minkowski spacetime}
Within the Clifford algebra $\C_n$ there exist eight automorphisms
\cite{Ras55,Var041} (including an identical automorphism $\Id$). We list
these transformations and their spinor representations:
\begin{eqnarray}
\cA\longrightarrow\cA^\star,&&\quad\sA^\star=\sW\sA\sW^{-1},\nonumber\\
\cA\longrightarrow\widetilde{\cA},&&\quad\widetilde{\sA}=\sE\sA^{\sT}\sE^{-1},
\nonumber\\
\cA\longrightarrow\widetilde{\cA^\star},&&\quad\widetilde{\sA^\star}=
\sC\sA^{\sT}\sC^{-1},\quad\sC=\sE\sW,\nonumber\\
\cA\longrightarrow\overline{\cA},&&\quad\overline{\sA}=\Pi\sA^\ast\Pi^{-1},
\nonumber\\
\cA\longrightarrow\overline{\cA^\star},&&\quad\overline{\sA^\star}=
\sK\sA^\ast\sK^{-1},\quad\sK=\Pi\sW,\nonumber\\
\cA\longrightarrow\overline{\widetilde{\cA}},&&\quad
\overline{\widetilde{\sA}}=\sS\left(\sA^{\sT}\right)^\ast\sS^{-1},\quad
\sS=\Pi\sE,\nonumber\\
\cA\longrightarrow\overline{\widetilde{\cA^\star}},&&\quad
\overline{\widetilde{\sA^\star}}=\sF\left(\sA^\ast\right)^{\sT}\sF^{-1},\quad
\sF=\Pi\sC.\nonumber
\end{eqnarray}
It is easy to verify that an automorphism set
$\{\Id,\,\star,\,\widetilde{\phantom{cc}},\,\widetilde{\star},\,
\overline{\phantom{cc}},\,\overline{\star},\,
\overline{\widetilde{\phantom{cc}}},\,\overline{\widetilde{\star}}\}$ of
$\C_n$ forms a finite group of order 8.
Let us give a physical interpretation of this
group.
\begin{sloppypar}
Let $\C_n$ be a Clifford algebra over the field $\F=\C$ and let
$\Ext(\C_n)=
\{\Id,\,\star,\,\widetilde{\phantom{cc}},\,\widetilde{\star},\,
\overline{\phantom{cc}},\,\overline{\star},\,
\overline{\widetilde{\phantom{cc}}},\,\overline{\widetilde{\star}}\}$
be an extended automorphism group\index{group!automorphism!extended}
of the algebra $\C_n$. Then there is
an isomorphism between $\Ext(\C_n)$ and a $CPT$--group
of the discrete transformations,
$\Ext(\C_n)\simeq\{1,\,P,\,T,\,PT,\,C,\,CP,\,CT,\,CPT\}\simeq
\dZ_2\otimes\dZ_2\otimes\dZ_2$. In this case, space inversion $P$, time
reversal $T$, full reflection $PT$, charge conjugation $C$, transformations
$CP$, $CT$ and the full $CPT$--transformation correspond to the automorphism
$\cA\rightarrow\cA^\star$, antiautomorphisms $\cA\rightarrow\widetilde{\cA}$,
$\cA\rightarrow\widetilde{\cA^\star}$, pseudoautomorphisms
$\cA\rightarrow\overline{\cA}$, $\cA\rightarrow\overline{\cA^\star}$,
pseudoantiautomorphisms $\cA\rightarrow\overline{\widetilde{\cA}}$ and
$\cA\rightarrow\overline{\widetilde{\cA^\star}}$, respectively.\end{sloppypar}
\begin{sloppypar}
The group $\{1,\,P,\,T,\,PT,\,C,\,CP,\,CT,\,CPT\}$ at the conditions
$P^2=T^2=(PT)^2=C^2=(CP)^2=(CT)^2=(CPT)^2=1$ and commutativity of all the
elements forms an Abelian group of order 8, which is isomorphic to a cyclic
group $\dZ_2\otimes\dZ_2\otimes\dZ_2$.
The multiplication table
of this group has a form\end{sloppypar}
\begin{center}{\renewcommand{\arraystretch}{1.4}
\begin{tabular}{|c||c|c|c|c|c|c|c|c|}\hline
     & $1$  & $P$  & $T$  & $PT$ & $C$  & $CP$ & $CT$ & $CPT$ \\ \hline\hline
$1$  & $1$  & $P$  & $T$  & $PT$ & $C$  & $CP$ & $CT$ & $CPT$ \\ \hline
$P$  & $P$  & $1$  & $PT$ & $T$  & $CP$ & $C$  & $CPT$& $CT$\\ \hline
$T$  & $T$  & $PT$ & $1$  & $P$  & $CT$ & $CPT$& $C$  & $CP$\\ \hline
$PT$ & $PT$ & $T$  & $P$  & $1$  & $CPT$& $CT$ & $CP$ & $C$\\ \hline
$C$  & $C$  & $CP$ & $CT$ & $CPT$& $1$  & $P$  & $T$  & $PT$\\ \hline
$CP$ & $CP$ & $C$  & $CPT$& $CT$ & $P$  & $1$  & $PT$ & $T$\\ \hline
$CT$ & $CT$ & $CPT$& $C$  & $CP$ & $T$  & $PT$ & $1$  & $P$\\ \hline
$CPT$& $CPT$& $CT$ & $CP$ & $C$  & $PT$ & $T$  & $P$  & $1$\\ \hline
\end{tabular}.
}
\end{center}
In turn, for the extended automorphism group
$\{\Id,\,\star,\,\widetilde{\phantom{cc}},\,\widetilde{\star},\,
\overline{\phantom{cc}},\,\overline{\star},\,
\overline{\widetilde{\phantom{cc}}},\,\overline{\widetilde{\star}}\}$
in virtue of commutativity $\widetilde{\left(\cA^\star\right)}=
\left(\widetilde{\cA}\right)^\star$,
$\overline{\left(\cA^\star\right)}=\left(\overline{\cA}\right)^\star$,
$\overline{\left(\widetilde{\cA}\right)}=
\widetilde{\left(\overline{\cA}\right)}$,
$\overline{\left(\widetilde{\cA^\star}\right)}=
\widetilde{\left(\overline{\cA}\right)^\star}$ and an involution property
$\star\star=\widetilde{\phantom{cc}}\widetilde{\phantom{cc}}=
\overline{\phantom{cc}}\;\overline{\phantom{cc}}=\Id$ we have a following
multiplication table
\begin{center}{\renewcommand{\arraystretch}{1.4}
\begin{tabular}{|c||c|c|c|c|c|c|c|c|}\hline
  & $\Id$ & $\star$ & $\widetilde{\phantom{cc}}$ & $\widetilde{\star}$ &
$\overline{\phantom{cc}}$ & $\overline{\star}$ &
$\overline{\widetilde{\phantom{cc}}}$ &
$\overline{\widetilde{\star}}$ \\ \hline\hline
$\Id$ & $\Id$ & $\star$ & $\widetilde{\phantom{cc}}$ & $\widetilde{\star}$ &
$\overline{\phantom{cc}}$ & $\overline{\star}$ &
$\overline{\widetilde{\phantom{cc}}}$ &
$\overline{\widetilde{\star}}$ \\ \hline
$\star$ & $\star$ & $\Id$ & $\widetilde{\star}$ & $\widetilde{\phantom{cc}}$ &
$\overline{\star}$ & $\overline{\phantom{cc}}$ &
$\overline{\widetilde{\star}}$ & $\overline{\widetilde{\phantom{cc}}}$\\ \hline
$\widetilde{\phantom{cc}}$ &
$\widetilde{\phantom{cc}}$ & $\overline{\star}$ & $\Id$ &
$\star$ & $\overline{\widetilde{\phantom{cc}}}$ & $\overline{\widetilde{\star}}$ &
$\overline{\phantom{cc}}$ & $\overline{\star}$\\ \hline
$\widetilde{\star}$ & $\widetilde{\star}$ & $\widetilde{\phantom{cc}}$ &
$\star$ & $\Id$ & $\overline{\widetilde{\star}}$ &
$\overline{\widetilde{\phantom{cc}}}$ &
$\overline{\star}$ & $\overline{\phantom{cc}}$\\ \hline
$\overline{\phantom{cc}}$ & $\overline{\phantom{cc}}$ & $\overline{\star}$ &
$\overline{\widetilde{\phantom{cc}}}$ & $\overline{\widetilde{\star}}$ & $\Id$ &
$\star$ & $\widetilde{\phantom{cc}}$ & $\widetilde{\star}$\\ \hline
$\overline{\star}$ & $\overline{\star}$ & $\overline{\phantom{cc}}$ &
$\overline{\widetilde{\star}}$ &
$\overline{\widetilde{\phantom{cc}}}$ & $\star$ &
$\Id$ & $\widetilde{\star}$ & $\widetilde{\phantom{cc}}$\\ \hline
$\overline{\widetilde{\phantom{cc}}}$ &
$\overline{\widetilde{\phantom{cc}}}$ &
$\overline{\widetilde{\star}}$ &
$\overline{\phantom{cc}}$ & $\overline{\star}$ &
$\widetilde{\phantom{cc}}$ & $\widetilde{\star}$ & $\Id$ & $\star$\\ \hline
$\overline{\widetilde{\star}}$ & $\overline{\widetilde{\star}}$ &
$\overline{\widetilde{\phantom{cc}}}$ & $\overline{\star}$ &
$\overline{\phantom{cc}}$ &
$\widetilde{\star}$ & $\widetilde{\phantom{cc}}$ & $\star$ & $\Id$\\ \hline
\end{tabular}.
}
\end{center}
The identity of multiplication tables proves the group isomorphism
\[
\{1,\,P,\,T,\,PT,\,C,\,CP,\,CT,\,CPT\}\simeq
\{\Id,\,\star,\,\widetilde{\phantom{cc}},\,\widetilde{\star},\,
\overline{\phantom{cc}},\,\overline{\star},\,
\overline{\widetilde{\phantom{cc}}},\,\overline{\widetilde{\star}}\}\simeq
\dZ_2\otimes\dZ_2\otimes\dZ_2.
\]
It has been recently shown \cite{Var041} that there exist 64 universal
coverings of the orthogonal group $O(p,q)$:
\[
\pin^{a,b,c,d,e,f,g}(p,q)\simeq\frac{(\spin_+(p,q)\odot
C^{a,b,c,d,e,f,g})}{\dZ_2},
\]
where
\[
C^{a,b,c,d,e,f,g}=\{\pm 1,\,\pm P,\,\pm T,\,\pm PT,\,\pm C,\,\pm CP,\,
\pm CT,\,\pm CPT\}
\]
is {\it a full $CPT$-group}. $C^{a,b,c,d,e,f,g}$ is a finite group of
order 16 (a complete classification of these groups is given in \cite{Var041}).
At this point, the group
\[
\Ext(\cl_{p,q})=\frac{C^{a,b,c,d,e,f,g}}{\dZ_2}
\]
we will call also as {\it a generating group}.

Let us define a $CPT$-group for the Dirac field in $\R^{1,3}$
(Minkowski spacetime). As is known,
the famous Dirac equation
in the $\gamma$--basis looks like
\begin{equation}\label{Diraceq}
\left(i\gamma_0\frac{\partial}{\partial x_0}+
i\boldsymbol{\gamma}\frac{\partial}{\partial\bx}-m\right)\psi(x_0,\bx)=0.
\end{equation}\begin{sloppypar}\noindent
The invariance of the Dirac equation with respect to $P$--, $T$--, and
$C$--transformations leads to the following representation
(see, for example, \cite{BLP89} and also many other textbooks on quantum
field theory):\end{sloppypar}
\begin{equation}\label{PTC}
P\sim\gamma_0,\quad T\sim\gamma_1\gamma_3,\quad C\sim\gamma_2\gamma_0.
\end{equation}
Thus, we can form a finite group of order 8
\begin{multline}
\{1,\,P,\,T,\,PT,\,C,\,CP,\,CT,\,CPT\}\sim\\
\sim\left\{1,\,\gamma_0,\,\gamma_1\gamma_3,\,\gamma_0\gamma_1\gamma_3,\,
\gamma_2\gamma_0,\,\gamma_2,\,\gamma_2\gamma_0\gamma_1\gamma_3,\,
\gamma_2\gamma_1\gamma_3\right\}.
\label{DirG2}
\end{multline}
The latter group should be understood as a generating group for the
$CPT$ group of the Dirac field in $\R^{1,3}$.
It is easy to verify that a multiplication table
of this group has a form
\begin{center}{\renewcommand{\arraystretch}{1.4}
\begin{tabular}{|c||c|c|c|c|c|c|c|c|}\hline
  & $1$ & $\gamma_0$ & $\gamma_{13}$ & $\gamma_{013}$ & $\gamma_{20}$ &
$\gamma_2$ & $\gamma_{2013}$ & $\gamma_{213}$\\ \hline\hline
$1$  & $1$ & $\gamma_0$ & $\gamma_{13}$ & $\gamma_{013}$ & $\gamma_2$ &
$\gamma_2$ & $\gamma_{2013}$ & $\gamma_{213}$\\ \hline
$\gamma_0$ & $\gamma_0$ & $1$ & $\gamma_{013}$ & $\gamma_{13}$ & $-\gamma_2$ &
$-\gamma_{20}$ & $-\gamma_{213}$ & $-\gamma_{2013}$\\ \hline
$\gamma_{13}$ & $\gamma_{13}$ & $\gamma_{013}$ & $-1$ & $-\gamma_0$ &
$\gamma_{2013}$ & $\gamma_{213}$ & $-\gamma_{20}$ & $-\gamma_2$\\ \hline
$\gamma_{013}$ & $\gamma_{013}$ & $\gamma_{13}$ & $-\gamma_0$ &
 $-1$ & $-\gamma_{213}$ & $-\gamma_{2013}$ & $\gamma_2$ &
$\gamma_{20}$\\ \hline
$\gamma_{20}$ & $\gamma_{20}$ & $\gamma_2$ & $\gamma_{2013}$ &
$\gamma_{213}$ & $1$ & $\gamma_0$ & $\gamma_{13}$ &
$\gamma_{013}$\\ \hline
$\gamma_2$ & $\gamma_2$ & $\gamma_{20}$ & $\gamma_{213}$ & $\gamma_{2013}$ &
$-\gamma_0$ & $-1$ & $-\gamma_{013}$ & $-\gamma_{13}$\\ \hline
$\gamma_{2013}$ & $\gamma_{2013}$ & $\gamma_{213}$ & $-\gamma_{20}$ &
$\gamma_2$ & $\gamma_{13}$ & $\gamma_{013}$ & $-1$ & $-\gamma_0$\\ \hline
$\gamma_{213}$ & $\gamma_{213}$ & $\gamma_{2013}$ & $-\gamma_2$ &
$-\gamma_{20}$ & $-\gamma_{013}$ & $-\gamma_{13}$ & $\gamma_0$ & $1$\\ \hline
\end{tabular}.
}
\end{center}
\begin{sloppypar}
Hence it follows that the group (\ref{DirG2}) is a non--Abelian
finite group of the order structure (3,4). In more details, it is the group
$\overset{\ast}{\dZ}_4\otimes\dZ_2$ with the signature
$(+,-,-,+,-,-,+)$.\end{sloppypar}
Therefore, the $CPT$-group in $\R^{1,3}$ is
\[
C^{+,-,-,+,-,-,+}\simeq\overset{\ast}{\dZ}_4\otimes\dZ_2\otimes\dZ_2.
\]
$C^{+,-,-,+,-,-,+}$ is a subgroup of the spacetime group $G(1,3)$.
The universal covering of the general Lorentz group is defined as
\begin{eqnarray}
\pin^{+,-,-,+,-,-,+}(1,3)&\simeq&\frac{(\spin_+(1,3)\odot
\overset{\ast}{\dZ}_4\otimes\dZ_2\otimes\dZ_2)}{\dZ_2},\nonumber\\
&\simeq&\frac{(SL(2,\C)\odot\overset{\ast}{\dZ}_4\otimes\dZ_2\otimes\dZ_2)}
{\dZ_2}.\nonumber
\end{eqnarray}
Instead of approximate equalities (\ref{PTC}) we can take exact equalities
of the two types considered recently by Socolovsky \cite{Soc04}:
$P=i\gamma_0$, $T=i\gamma_{13}$, $C=\gamma_{02}$ and
$P=i\gamma_0$, $T=\gamma_{13}$, $C=i\gamma_{02}$. It is easy to verify
that the equalities of the second type lead to the group
$C^{+,-,-,+,-,-,+}$.

Let us study an extended automorphism group of the spacetime algebra
$\cl_{1,3}$.
Using the matrices of the canonical basis
(\ref{CB}), we define elements
of the group $\sExt(\cl_{1,3})$. First of all, the matrix of the automorphism
$\cA\rightarrow\cA^\star$ has the form $\sW=\gamma_0\gamma_1\gamma_2\gamma_3$.
Further, since
\[
\gamma^{\sT}_0=\gamma_0,\quad\gamma^{\sT}_1=-\gamma_1,\quad
\gamma^{\sT}_2=-\gamma_2,\quad\gamma^{\sT}_3=-\gamma_3,
\]
then in accordance with $\widetilde{\sA}=\sE\sA^{\sT}\sE^{-1}$ we have
\[
\gamma_0=\sE\gamma_0\sE^{-1},\quad\gamma_1=-\sE\gamma_1\sE^{-1},\quad
\gamma_2=\sE\gamma_2\sE^{-1},\quad\gamma_3=-\sE\gamma_3\sE^{-1}.
\]
Hence it follows that $\sE$ commutes with $\gamma_0$, $\gamma_2$ and
anticommutes with $\gamma_1$, $\gamma_3$, that is,
$\sE=\gamma_1\gamma_3$. From the definition $\sC=\sE\sW$ we find that
the matrix of the antiautomorphism $\cA\rightarrow\widetilde{\cA^\star}$
has the form $\sC=\gamma_0\gamma_2$.
The canonical $\gamma$--basis contains both complex and real matrices:
\[
\gamma^\ast_0=\gamma_0,\quad\gamma^\ast_1=\gamma_1,\quad
\gamma^\ast_2=-\gamma_2,\quad\gamma^\ast_3=\gamma_3.
\]
Therefore, from $\overline{\sA}=\Pi\sA^\ast\Pi^{-1}$ we have
\[
\gamma_0=\Pi\gamma_0\Pi^{-1},\quad\gamma_1=\Pi\gamma_1\Pi^{-1},\quad
\gamma_2=-\Pi\gamma_2\Pi^{-1},\quad\gamma_3=\Pi\gamma_3\Pi^{-1}.
\]
From the latter relations we obtain
$\Pi=\gamma_0\gamma_1\gamma_3$. Further,
in accordance with $\sK=\Pi\sW$ for the matrix of the pseudoautomorphism
$\cA\rightarrow\overline{\cA^\star}$ we have $\sK=\gamma_2$. Finally,
for the pseudoantiautomorphisms
$\cA\rightarrow\overline{\widetilde{\cA}}$,
$\cA\rightarrow\overline{\widetilde{\cA^\star}}$ from the definitions
$\sS=\Pi\sE$, $\sF=\Pi\sC$ it follows $\sS=\gamma_0$,
$\sF=\gamma_1\gamma_2\gamma_3$. Thus, we come to the following extended
automorphism group:
\begin{multline}
\sExt(\cl_{1,3})\simeq\{\sI,\,\sW,\,\sE,\,\sC,\,\Pi,\,\sK,\,\sS,\,\sF\}\simeq\\
\{\sI,\,\omega=\gamma_0\gamma_1\gamma_2\gamma_3,\,\gamma_1\gamma_3,\,
\gamma_0\gamma_2,\,\gamma_0\gamma_1\gamma_3,\,\gamma_2,\,\gamma_0,\,
\gamma_1\gamma_2\gamma_3\}.\label{Dirac3}
\end{multline}
The multiplication table of this group has a form:
\begin{center}{\renewcommand{\arraystretch}{1.4}
\begin{tabular}{|c||c|c|c|c|c|c|c|c|}\hline
  & $\sI$ & $\omega$ & $\gamma_{13}$ & $\gamma_{02}$ & $\gamma_{013}$ &
$\gamma_{2}$ & $\gamma_{0}$ & $\gamma_{123}$\\ \hline\hline
$\sI$  & $\sI$ & $\omega$ & $\gamma_{13}$ & $\gamma_{02}$ & $\gamma_{013}$ &
$\gamma_{2}$ & $\gamma_{0}$ & $\gamma_{123}$\\ \hline
$\omega$ & $\omega$ & $-\sI$ & $\gamma_{02}$ & $-\gamma_{012}$
& $-\gamma_2$ &
$\gamma_{013}$ & $-\gamma_{123}$ & $\gamma_{0}$\\ \hline
$\gamma_{13}$ & $\gamma_{13}$ & $\gamma_{02}$ & $-\sI$ & $-\omega$ &
$-\gamma_{0}$ & $-\gamma_{123}$ & $\gamma_{013}$ & $\gamma_2$\\ \hline
$\gamma_{02}$ & $\gamma_{02}$ & $-\gamma_{13}$ & $-\omega$ &
$\sI$ & $\gamma_{123}$ & $-\gamma_{0}$ & $-\gamma_2$ &
$\gamma_{013}$\\ \hline
$\gamma_{013}$ & $\gamma_{013}$ & $\gamma_2$ & $-\gamma_{0}$ &
$-\gamma_{123}$ & $-\sI$ & $-\omega$ & $\gamma_{13}$ &
$\gamma_{02}$\\ \hline
$\gamma_2$ & $\gamma_2$ & $-\gamma_{013}$ & $-\gamma_{123}$ & $\gamma_{0}$ &
$\omega$ & $-\sI$ & $-\gamma_{02}$ & $\gamma_{13}$\\ \hline
$\gamma_{0}$ & $\gamma_{0}$ & $\gamma_{123}$ & $\gamma_{013}$ &
$\gamma_2$ & $\gamma_{13}$ & $\gamma_{02}$ & $\sI$ & $\omega$\\ \hline
$\gamma_{123}$ & $\gamma_{123}$ & $-\gamma_{0}$ & $\gamma_2$ &
$-\gamma_{013}$ & $-\gamma_{02}$ & $\gamma_{13}$ & $-\omega$ &
$\sI$\\ \hline
\end{tabular}.
}
\end{center}
As follows from this table, the group $\sExt(\cl_{1,3})$ is non-Abelian.
$\sExt(\cl_{1,3})$ contains Abelian group of spacetime reflections
$\sAut_-(\cl_{1,3})\simeq\dZ_4$ as a subgroup.
More precisely, the group (\ref{Dirac3}) is a finite group
$\overset{\ast}{\dZ}_4\otimes\dZ_2$ with the signature
$(-,-,+,-,-,+,+)$.

We see that the generating groups (\ref{DirG2}) and
(\ref{Dirac3}) are isomorphic:
\[
\{1,\,P,\,T,\,PT,\,C,\,CP,\,CT,\,CPT\}\simeq
\{\sI,\,\sW,\,\sE,\,\sC,\,\Pi,\,\sK,\,\sS,\,\sF\}\simeq
\overset{\ast}{\dZ}_4\otimes\dZ_2.
\]
Moreover, the subgroups of spacetime reflections of these groups are
also isomorphic:
\[
\{1,\,P,\,T,\,PT\}\simeq\{\sI,\,\sW,\sE,\,\sC\}\simeq\dZ_4.
\]
Thus, we come to the following result: the finite group (\ref{DirG2}),
derived from the analysis of invariance properties of the Dirac equation
with respect to discrete transformations $C$, $P$ and $T$, is isomorphic
to an extended automorphism group of the Dirac algebra $\C_4$.
This result allows us to study discrete symmetries and their group
structure for physical fields of any spin (without handling to analysis
of relativistic wave equations).

\section{The $CPT$-group in the de Sitter space}
Let us define now a $CPT$-group of the Dirac field in $\R^{4,1}$.
First of all, using the spinbasis (\ref{SS})
we construct the generating group
$\sExt(\cl_{4,1})=\{\sI,\sW,\sE,\sC,\Pi,\sK,\sS,\sF\}$.
It is obvious that for the matrix of the automorphism
$\cA\rightarrow\cA^\star$ we have $\sW=\gamma_{12345}$. With the view to
find a spinor representation for the antiautomorphism
$\cA\rightarrow\widetilde{\cA}$ we see that
\[
\gamma^{\sT}_1=\gamma_1,\quad\gamma^{\sT}_2=\gamma_2,\quad
\gamma^{\sT}_3=-\gamma_3,\quad\gamma^{\sT}_4=-\gamma_4,\quad
\gamma^{\sT}_5=\gamma_5.
\]
From $\widetilde{\sA}=\sE\sA^{\sT}\sE^{-1}$ it follows
\begin{gather}
\gamma_1=\sE\gamma_1\sE^{-1},\quad\gamma_2=\sE\gamma_2\sE^{-1},\quad
\gamma_3=-\sE\gamma_3\sE^{-1},\nonumber\\
\gamma_4=-\sE\gamma_4\sE^{-1},\quad\gamma_5=\sE\gamma_5\sE^{-1}.\nonumber
\end{gather}
Therefore, the matrix $\sE$ of the transformation
$\cA\rightarrow\widetilde{\cA}$ commutes with $\gamma_1$, $\gamma_2$,
$\gamma_5$ and anticommutes with $\gamma_3$, $\gamma_4$.
It is easy to verify that $\sE=\gamma_{34}$. From $\sC=\sE\sW^{\sT}$ we
obtain $\sC=\gamma_{125}$ for the matrix of the antiautomorphism
$\cA\rightarrow\widetilde{\cA^\star}$.

We see that in the spinbasis of $\cl_{4,1}$ there are both complex and
real matrices:
\[
\gamma^\ast_1=\gamma_1,\quad\gamma^\ast_2=\gamma_2,\quad
\gamma^\ast_3=-\gamma_3,\quad\gamma^\ast_4=-\gamma_4,\quad
\gamma^\ast_5=-\gamma_5.
\]
Therefore, from $\overline{\sA}=\Pi\sA^\ast\Pi^{-1}$ it follows that
\begin{gather}
\gamma_1=\Pi\gamma_1\Pi^{-1},\quad\gamma_2=\Pi\gamma_2\Pi^{-1},\quad
\gamma_3=-\Pi\gamma_3\Pi^{-1},\nonumber\\
\quad\gamma_4=-\Pi\gamma_4\Pi^{-1},\quad
\gamma_5=-\Pi\gamma_5\Pi^{-1}.\nonumber
\end{gather}
From the latter relations we find $\Pi=\gamma_{123}$ for the
pseudoautomorphism $\cA\rightarrow\overline{\cA}$. Further, using the
definitions $\sK=\Pi\sW$, $\sS=\Pi\sE$ and $\sF=\Pi\sC$ we have
$\sK=\gamma_{45}$, $\sS=\gamma_{124}$ and $\sF=\gamma_{35}$ for the
transformations $\cA\rightarrow\overline{\cA^\star}$,
$\cA\rightarrow\overline{\widetilde{\cA}}$ and
$\cA\rightarrow\overline{\widetilde{\cA^\star}}$. Thus,
\[
\sExt(\cl_{4,1})=\{\sI,\omega=\gamma_{12345},\gamma_{34},
\gamma_{125},\gamma_{123},
\gamma_{45},\gamma_{124},\gamma_{35}\}.
\]
The multiplication table of this group has the form
\begin{center}{\renewcommand{\arraystretch}{1.4}
\begin{tabular}{|c||c|c|c|c|c|c|c|c|}\hline
  & $\sI$ & $\omega$ & $\gamma_{34}$ & $\gamma_{125}$ & $\gamma_{123}$ &
$\gamma_{45}$ & $\gamma_{124}$ & $\gamma_{35}$\\ \hline\hline
$\sI$  & $\sI$ & $\omega$ & $\gamma_{34}$ & $\gamma_{125}$ & $\gamma_{123}$ &
$\gamma_{45}$ & $\gamma_{124}$ & $\gamma_{35}$\\ \hline
$\omega$ & $\omega$ & $-\sI$ & $-\gamma_{125}$ & $\gamma_{34}$
& $-\gamma_{45}$ &
$\gamma_{123}$ & $\gamma_{35}$ & $-\gamma_{124}$\\ \hline
$\gamma_{34}$ & $\gamma_{34}$ & $-\gamma_{125}$ & $-\sI$ & $\omega$ &
$-\gamma_{124}$ & $\gamma_{35}$ & $\gamma_{123}$ & $-\gamma_{45}$\\ \hline
$\gamma_{125}$ & $\gamma_{125}$ & $\gamma_{34}$ & $\omega$ &
$\sI$ & $\gamma_{35}$ & $\gamma_{124}$ & $\gamma_{45}$ &
$-\gamma_{123}$\\ \hline
$\gamma_{123}$ & $\gamma_{123}$ & $-\gamma_{45}$ & $\gamma_{124}$ &
$-\gamma_{35}$ & $-\sI$ & $\omega$ & $-\gamma_{34}$ &
$\gamma_{125}$\\ \hline
$\gamma_{45}$ & $\gamma_{45}$ & $\gamma_{123}$ & $-\gamma_{35}$ & $-\gamma_{124}$ &
$\omega$ & $\sI$ & $-\gamma_{125}$ & $\gamma_{34}$\\ \hline
$\gamma_{124}$ & $\gamma_{124}$ & $\gamma_{35}$ & $-\gamma_{123}$ &
$\gamma_{45}$ & $-\gamma_{34}$ & $\gamma_{125}$ & $-\sI$ & $-\omega$\\ \hline
$\gamma_{35}$ & $\gamma_{35}$ & $-\gamma_{124}$ & $\gamma_{45}$ &
$-\gamma_{123}$ & $-\gamma_{125}$ & $\gamma_{34}$ & $-\omega$ &
$\sI$\\ \hline
\end{tabular}.
}
\end{center}
From the table we see that
\[
\sExt(\cl_{4,1})\simeq\overset{\ast}{\dZ}_4\otimes\dZ_2.
\]
Therefore, an universal covering of the de Sitter group $O(4,1)$ is
\[
\pin^{-,-,+,-,+,-,+}(4,1)\simeq\frac{(\spin_+(4,1)\odot
C^{-,-,+,-,+,-,+})}{\dZ_2},
\]
where
\[
C^{-,-,+,-,+,-,+}\simeq\overset{\ast}{\dZ}_4\otimes\dZ_2\otimes\dZ_2
\]
is {\it a full $CPT$ group of the Dirac field in the space $\R^{4,1}$}.
In turn, $C^{-,-,+,-,+,-,+}$ is a subgroup of the Dirac group $G(4,1)$.%


\vskip 30pt
\begin{thebibliography}{00}

\bibitem{BLP89} V. B. Berestetskii, E. M. Lifshitz, L.P. Pitaevskii,
{\em Quantum Electrodynamics}, Course of Theoretical Physics Vol.~4, 2nd Edition
(Pergamon Press, Oxford, 1982).
\bibitem{Bra85} H. W. Braden, ``$N$-dimensional spinors: Their properties
in terms of finite groups," J. Math. Phys. {\bf 26}, 613--620 (1985).
\bibitem{BW35} R. Brauer, H. Weyl, ``Spinors in $n$ dimensions,"
Amer. J. Math. {\bf 57}, 425-449 (1935).
\bibitem{Che54} C. Chevalley, {\em The Algebraic Theory of Spinors}
(Columbia University Press, New York, 1954).
\bibitem{Dab88} L. D\c{a}browski, {\em Group Actions on Spinors}
(Bibliopolis, Naples, 1988).
\bibitem{Dir35} P. A. M. Dirac, ``The electron wave equation in de Sitter
space," Annals of Math. {\bf 36}, 657 (1935).
\bibitem{FK69} W. L. Fushchych, I. Yu. Krivsky, ``On representations of the
inhomogeneous de Sitter group and equations in five-dimensional Minkowski
space," Nucl. Phys. {\bf B14}, 573--585 (1969).
\bibitem{LW66} T. D. Lee, G. C. Wick, ``Space inversion, time reversal,
and other discrete symmetries," Phys. Rev. {\bf 148}, 1385--1404 (1966).
\bibitem{Ras55} P. K. Rashevskii, ``The Theory of Spinors," (in Russian)
Uspekhi Mat. Nauk {\bf 10}, 3--110 (1955); English translation in
Amer. Math. Soc. Transl. (Ser.~2) {\bf 6}, 1 (1957).
\bibitem{Sal81a} N. Salingaros, ``Realization, extension, and
classification of certain physically important groups and algebras,"
J. Math. Phys. {\bf 22}, 226--232, (1981).
\bibitem{Sal82} N. Salingaros, ``On the classification of Clifford
algebras and their relation to spinors in $n$ dimensions," J. Math. Phys.
{\bf 23}(1), (1982).
\bibitem{Sal84} N. Salingaros, ``The relationship between finite
groups and Clifford algebras," J. Math. Phys. {\bf 25}, 738--742, (1984).
\bibitem{Sha94} R. Shaw, ``Finite geometry, Dirac groups and the table
of real Clifford algebras," Univ. of Hull Maths Research Report, {\bf 7}, no 1,
(1994).
\bibitem{Shi60} Yu. M. Shirokov, ``Spacial and time reflections in the
relativistic theory," Zh. Ehksp. Teor. Fiz. {\bf 38}, 140--150 (1960).
\bibitem{Soc04} M. Socolovsky, ``The CPT group of the Dirac field,"
Int. J. Theor. Phys. {\bf 43}, 1941--1967 (2004).
\bibitem{Var99} V. V. Varlamov, ``Fundamental Automorphisms of
Clifford Algebras and an Extension of D\c{a}browski Pin Groups,"
Hadronic J. {\bf 22}, 497--535 (1999).
\bibitem{Var00} V. V. Varlamov, ``Discrete Symmetries and Clifford Algebras,"
Int. J. Theor. Phys. {\bf 40}, No. 4, 769--805 (2001).
\bibitem{Var041} V. V. Varlamov, ``Group Theoretical Interpretation
of the $CPT$-theorem," in
{\em Mathematical Physics Research at the Cutting Edge} (Ed. C. V. Benton),
pp. 51--100 (Nova Science Publishers, New York, 2004).
\bibitem{Var042} V. V. Varlamov, ``Universal Coverings of the Orthogonal
Groups," Advances in Applied Clifford Algebras {\bf 14}(1), 81--168
(2004).
\end{thebibliography}
\end{document}